\documentstyle[aps,preprint]{revtex}

\begin{document}
\draft
\title{Multi-shell gold nanowires under compression}
\tightenlines
\author{G. Bilalbegovi\'c}
\address{Department of Physics, University of Rijeka, Omladinska 14,51000 Rijeka,\\
Croatia}
\date{to be published in J. Phys: Condens. Matter}
\maketitle

\begin{abstract}
Deformation properties of multi-wall gold nanowires under compressive
loading are studied. Nanowires are simulated using a realistic many-body
potential. Simulations start from cylindrical fcc (111) structures at $T=0$
K. After annealing cycles axial compression is applied on multi-shell
nanowires for a number of radii and lengths at $T=300$ K. Several types of
deformation are found, such as large buckling distortions and progressive
crushing. Compressed nanowires are found to recover their initial lengths
and radii even after severe structural deformations. However, in contrast to
carbon nanotubes irreversible local atomic rearrangements occur even under
small compressions.
\end{abstract}

\section{Introduction}

Carbon nanotubes are the subject of intensive research last several years 
\cite{Phyworld}. Their electrical and thermal conductivity, as well as
mechanical properties are of great scientific and technological interest.
Depending on the diameter and the helicity carbon nanotubes can be either
metallic or semiconducting, and exhibit quantum wire properties. Therefore,
they are important for nanoelectronic devices. Promising applications of
carbon nanotubes are also based on their extraordinary mechanical
properties. Investigations have shown that carbon nanotubes are among
strongest materials \cite{Phyworld}. The effects of deformations and various
defects are studied for single-wall and multi-wall tubes, as well as for
ropes composed of carbon nanotubes \cite
{Avouris,Yakobson,Cornwell,Srivastava,Ozaki}. It was found that very
distorted carbon nanotubes return to their original form when loading is
released. As a consequence, carbon nanotubes are useful in applications,
such as high-strength fibers, components of composite materials and tips in
scanning-probe microscopy \cite
{Phyworld,Avouris,Yakobson,Cornwell,Srivastava,Ozaki}.

The property of carbon nanotubes to deform elastically is in part the
outcome of the strength and rigidity of the graphitic bond. The cylindrical
structure of nanotubes increases their elasticity and strength. It is
important to investigate mechanical properties for cylindrical
nanostructures made of other materials where a different type of bonding
exists. These investigations may shed light onto the choice of optimal
materials and structures for elements in nanomechanical devices. Metallic
wires are important for applications. For example, gold wires are already
used as interconnections in the chips. The miniaturization of electronic and
mechanical devices requires investigations of very small wires whose
diameters and lengths are in the range from $1$ nm to $10$ nm. Recently a
technique for fabrication of metallic nanowires with diameters down to $d=3$
nm was invented \cite{Natelson}. This method is based on the thickness
resolution of molecular-beam epitaxy and formation of extremely precise
templates for metal deposition. The mechanical properties of metallic
nanowires till now were investigated because of the experiments dealing with
the contact between two pieces of metal. For example, metallic nanobridges
were obtained in Scanning Tunneling Microscopy (STM) experiments \cite
{Pascual}, mechanically controllable break-junctions \cite{Muller}, and
macroscopic table top metallic contacts \cite{Costa}. In this context the
stability of tip-suspended nanowires, their behavior under tensile strain,
fracture properties, and conductance were investigated by Molecular Dynamics
(MD) simulations and density functional theory calculations \cite
{Brandbyge,Bratkovsky,Torres,Landman,Barnett,Ciraci,Finbow,Hakkinen,Taraschi,Nakamura,Ikeda,Branicio,Tosatti,Baolin}%
. Mechanical properties of metallic nanocontacts under compressive strain
are less studied \cite{Torres,Landman,Hakkinen}. However, compression of
free single-wall carbon nanotubes was investigated by MD simulations \cite
{Yakobson,Cornwell,Srivastava,Ozaki}.

Multi-shell structures were obtained in MD simulations of finite and
infinite gold nanowires \cite{PRB,MolSim,SSC}. Finite nanowires with radii
around a nanometer and of a length/diameter ratio between $1$ and $3$ were
studied \cite{PRB,MolSim}. Cylindrical multi-shell structures were found for
a length/diameter ratio between $2$ and $3$. Structures and melting of
infinite gold nanowires with radii around a nanometer and an initial
orientation along the (111), (110), and (100) directions were also
investigated \cite{SSC}. The results have shown that a formation of
cylindrical shells is the most pronounced for an initial fcc (111)
orientation. Similar gold nanostructures were found in experiments \cite
{Ohnishi,Yanson,Ugarte}. These studies are enabled by recent advances in
microscopic techniques which now produce images of atomic scale resolution.
Gold nanostructures were prepared by contacting a gold substrate with a STM
tip, and they were simultaneously imaged using an ultrahigh-vacuum electron
microscope \cite{Ohnishi,Ugarte}. The conductance of these gold
nanostructures was measured during retraction of the STM tip. It was found
that the conductance is quantized in units of $2e^{2}/h$, where $e$ is the
electron charge and $h$ is the Planck's constant \cite{Ohnishi,Yanson,Ugarte}%
. Stable and regular strands of gold atoms that are from one nanometer to
several nanometers in length were observed. The diameters of these wires
were around one nanometer. In MD simulations of Pb and Al ultrathin infinite
wires at $T=0$ K several exotic structures were found, for example
icosahedral and helical forms \cite{Gulseren}. Similar simulation has been
carried out for gold nanowires \cite{Baolin}. Aluminum and copper infinite
nanowires were also simulated at the room temperature and multi-shell and
filled structures were obtained \cite{CompMatSci}. Recently Zach and
coworkers have developed a method for producing metallic wires with
diameters ranging from $15$ nanometers to $1.0$ micrometers and lengths of
up to half a millimeter \cite{Zach}. They were using the graphite substrate
as a template for growing molybdenum wires. The mechanical strength of these
nanowires was tested and it was found that they were able sometimes to bend
at 90 degrees without breaking. In this work a MD simulation study of free
multi-shell gold nanowires under axial compression at $T=300$ K is
presented. Our results show that these gold nanostructures are able to
sustain very large values of the strain and recover initial radii and
lenghts even after severe deformations. In the following the interatomic
potential, the wires preparation method, and simulation details are
described in Sec. II. Results and discussion are presented in Sec. III. A
summary and conclusions are given in Sec. IV.

\section{Computational method}

The interaction between gold atoms was modeled by the well-tested
embedded-atom potential \cite{Furio}. It is known that embedded-atom
potentials provide a satisfactory description of metallic bonding in the
bulk, for surfaces, and nanoparticles \cite{Daw}. In comparison with pair
interactions, such as the Lennard-Jones and Morse potentials, the
embedded-atom model takes an account of many-body effects in metals. The
particular realization of the embedded-atom potential used here, the
so-called glue model, has been shown to accurately reproduce experimental
values for a wide range of physical properties of gold \cite{Furio}. The
classical MD simulation method was carried out. The equations of motion were
integrated using a time step of $\Delta t = 7.14 \times 10^{-15}$ s. In all
simulations presented in this work the temperature control was realized
through velocity rescaling of all active atoms.

It is known that bulk gold crystallizes in a face-centered cubic lattice.
Initially, for nanowires at $T=0$ K, atoms were arranged in a fcc structure
with the (111) direction parallel to the axis of the wire. The prepared
nanowires were approximately cylindrical in cross-sections. All atoms
further than a chosen value of the radius from the nanowire axis were
removed. Several nanowires with the length $L_0$ between $4$ nm and $12$ nm,
the number of atoms between $540$ and $2067$, and the radius $R_0$ between $%
0.5$ nm and $1.2$ nm were investigated. All wires were first relaxed at $T=0$
K. Then samples were heated to $1000$ K. This was followed by a quench to $%
T=0$ K, and heating to $T=300$ K. In previous simulations, where structural
and vibrational properties of finite gold nanowires were investigated, MD
boxes in the annealing cycles were heated to $600$ K \cite{PRB,MolSim}.
However, it was found here that even finite nanowires are robust, and
preserve their cylindrical structures when heated up to $\sim 0.75$ of the
bulk melting temperature. Properties of nanowires under compression were
investigated after evolution of $10^5$ time steps at $T=300$ K. As in
previous simulations for finite \cite{PRB,MolSim} and infinite gold
nanowires \cite{SSC}, after a preparation procedure stable multi-shell
cylindrical structures were obtained. In comparison with previous
simulations of finite gold nanowires, a higher annealing temperature applied
here produces more regular multi-shell structures. After equilibration, as
in simulations of single-wall carbon nanotubes \cite
{Yakobson,Cornwell,Srivastava,Ozaki}, the edge rings were fixed, and then
compressive axial loading was applied. These simulations under compression
most often were taken out to $2 \times 10^4$ time steps. It was checked that
typical deformation patterns formed during this time did not change up to $%
10^5$ time steps.

\section{Results and discussion}

Several structures of multi-shell wires at $300$ K and before compression
was applied are presented in Fig. 1. Figure 1(a) shows a three-shell
structure with the central core. This type of the multi-shell configuration
with a single central strand of gold atoms forms most often. A nanowire with
the smallest investigated radius forms a two-wall structure with a large
empty core shown in Fig. 1(b). The core of two strands of atoms, as in the
four-wall structure shown in Fig. 1(c), was also obtained.

Short and narrow nanowires (e.g., $R_{0}=0.9$ nm, $L_{0}=4$ nm) under
compression exhibit only one morphological pattern. They progressively
shorten when compression increases, but remain straight. As an exception,
long and extremely narrow nanowires (e.g., $R_{0}=0.5$ nm, $L_{0}=12$ nm)
under compression often crush into irregular ellipsoidal morphologies.
Thicker short nanowires (e.g., $R_{0}=1.2$ nm, $L_{0}=8$ nm) show three
morphological patterns. For small compressions these nanowires deform by
rippling. However, rippling is less pronounced than in single-wall carbon
nanotubes \cite{Ozaki}, and only isolated ripples appear. At $16$ GPa of the
stress one end of the nanowire deforms more than the other. Such a
deformation is shown in Fig. 2. For larger stress ($\geq 32$ GPa) these
nanowires also deform by crushing, flatten and preserve their cylindrical
shapes. Nanowires return to the configuration with approximately initial $%
R_{0}$ and $L_{0}$ when compression is released. The outer cylindrical
surface of those straight wires is more rough than in initial ones.

The most interesting behavior exhibit long and narrow nanowires [e.g., $%
R_{0}=0.9$ nm, $L_{0}=12$ nm, shown in Fig. 3(a)]. These slender nanowires
under compression buckle and sometimes exhibit largely distorted
configurations. Figure 3(b) shows deformation of the nanowire in Fig. 3(a)
at $4.81$ GPa of the stress. The structure formed in this sideways
deformation is almost symmetric. Figure 4(a) shows a configuration of the
same nanowire at $6.41$ GPa of the stress. Although, deformed substantially
[Fig. 4(a)], this structure immediately straightens [Fig. 4(b)] when
compression is released. This structural change already takes place within $%
7 $ ps of a time evolution after loading is released. The resulting
configuration (shown in Fig. 4(b) after $35$ ps, i.e., $5\times 10^{3}$ time
steps) is less regular than initial nanowire, but has the same average
radius and length. In further simulations at the same temperature this
nanowire does not change substantially up to $0.7$ ns. The same stress of $%
6.41$ GPa applied on the structure shown in Fig. 4(b) produces a buckled
morphology similar to one shown in Fig. 4(a). When loading is released in
this repeated compression, a nanowire again returns to the straight form
similar to one shown in Fig. 4(b). Large buckling deformations [as shown in
Figs. 3(b) and 4(a)] exist for average compressions. For the highest values
of the stress slender nanowires deform by crushing and flattening, and
remain straight.

These simulations show that a multi-shell structure disappears under
compression. The density plots in Fig. 5 illustrate the changes in a
multi-shell structure under compression. For the stress of $3.2$ GPa the
shell structure is still present [Fig. 5(b)], whereas at $4.81$ GPa the
shells are much less pronounced [Fig. 5(c)]. The shell structure is almost
absent at $6.41$ GPa [Fig. 5(d)]. When compression is released nanowires
straighten, but their internal multi-shell structure only partially recovers
in the simulation of $2\times 10^{4}$ time steps. Disordering of a
multi-shell structure is more pronounced in longer nanowires and for a
higher applied stress. As in initial annealing cycles, it is possible to
improve this partially ordered multi-shell structure by heating. In
multi-shell gold nanowires atomic rearrangements occur for the smallest
applied stress. In contrast, carbon nanotubes up to certain values of stress
are in the elastic regime. There virtually no defects are found after
compressive loading is released \cite{Yakobson,Cornwell,Srivastava,Ozaki}.
Beyond the elastic regime carbon nanotubes deform plastically. In the
plastic regime, as here for gold nanowires, compressed carbon nanotubes
recover from severe structural deformations, but local atomic rearrangements
exist.

A plot of the stress vs. the strain for several nanowires is presented in
Fig. 6. All shells of multi-wall nanowires are stressed and it is assumed
that the axial strain and stress are uniformly distributed over the cross
section. Therefore, in calculation of the stress the whole cross-sectional
area of the nanowire is used. These results show that multi-shell gold
nanowires are able to sustain a large compressive stress. Multi-shell gold
nanowires at large compressions behave in similar way as ropes of
single-wall carbon nanotubes under pressure \cite{Chesnokov}. These ropes of
nanotubes act as a mechanical energy storage. This was attributed to
crushing and flattening of the tube cross section. A flattening along the
wire axis was found in this simulation for gold nanowires. Those
morphological patterns correspond to the vertical portions of the curves in
Fig. 6. Therefore, because of the property to recover their shapes even for
large values of the strain, multi-shell gold nanowires also act as a
mechanical energy storage. Although the chemical bonding in carbon nanotubes
and gold nanowires is different, the packing effects at nanometre length
scales are also important. Regular axially symmetrical distribution of
carbon nanotubes in ropes and cylindrical gold shells in nanowires enable
them to store a mechanical energy.

\section{Conclusions}

Molecular dynamics simulation was carried out to study deformations of
multi-shell gold nanowires under axial compressive loading at $T=300$ K.
This simulation is based on a well-tested embedded-atom potential. It was
found that multi-shell nanowires are able to sustain large values of the
compressive strain. The most interesting behavior show long and narrow
nanowires where large buckling distortions without a failure are possible
when axial compression is applied and then released. The property of carbon
nanotubes to recover without a damage after large deformations was explained
as a consequence of the carbon bonding in the graphite layer and the
cylindrical structure of the tube. The simulation presented here shows that
multi-shell gold nanowires after large deformations recover their
cylindrical forms, as well as initial radii and lenghts. A similar result
(i.e, that large bending deformations without breaking are possible) was
recently obtained experimentally for molybdenum nanowires \cite{Zach}. In
contrast to carbon nanotubes, because of different type of chemical bonding
in gold, atomic rearrangements occur even under small compressions. As a
result, defects and local structural changes are always present in gold
nanowires when loading is released. The property of multi-shell gold
nanowires to store a mechanical energy is useful for applications.
Mechanical properties of various metallic nanowires deserve further
experimental and theoretical investigations.

\acknowledgments

This work has been carried under the HR-MZT project 119206 ``Dynamical
Properties of Surfaces and Nanostructures'' and the EC Research Action COST
P3 ``Simulation of Physical Phenomena in Technological Applications''.

\clearpage

\begin{figure}[tbp]
\caption{ Top view of initial uncompressed multi-shell nanowires: (a) $%
R_{0}=0.9$ nm, $L_{0}=12$ nm, (b) $R_{0}=0.5$ nm, $L_{0}=12$ nm, (c) $%
R_{0}=1.2$ nm, $L_{0}=8.0$ nm. The trajectory plots refer to a time span of $%
\sim 7$ ps, and include all atoms in the slice of the thickness of $4$ nm
along the wire axis.}
\label{fig1}
\end{figure}

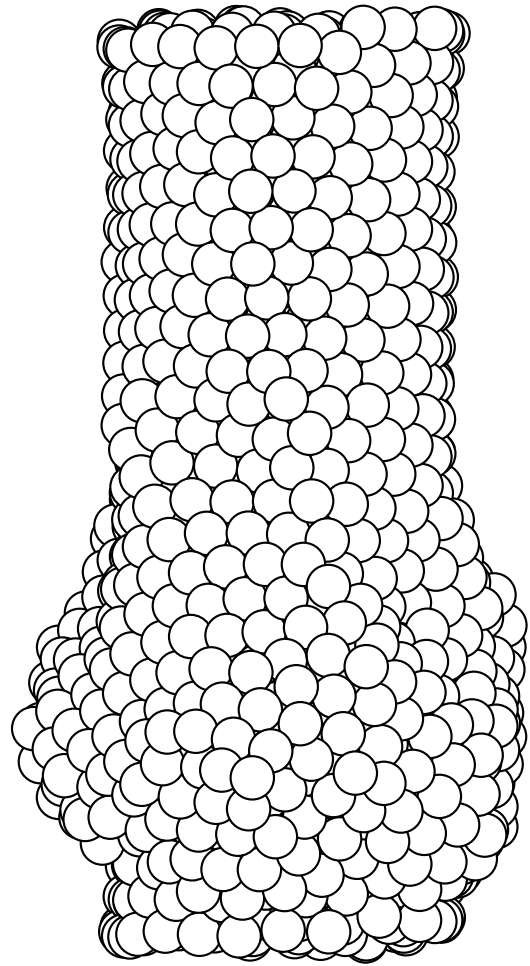
\begin{figure}[tbp]
\caption{Deformation of a nanowire with $R_0=1.2$ nm and $L_0=8$ nm, at the
stress of $19.23$ GPa.}
\label{fig2}
\end{figure}

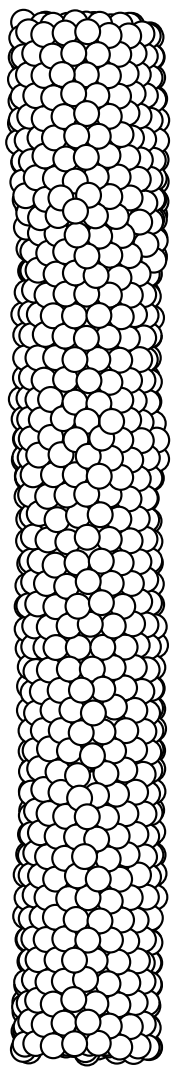
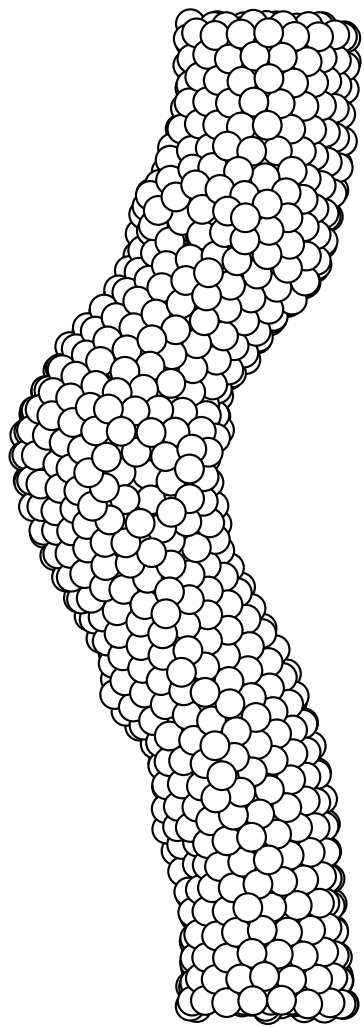
\begin{figure}[tbp]
\caption{ Nanowire with $R_0=0.9$ nm and $L_0=12$ nm: (a) uncompressed
configuration [a top view of the same nanowire is shown in Fig. 1(a)], (b)
deformation of this nanowire at the stress of $4.81$ GPa.}
\label{fig3}
\end{figure}

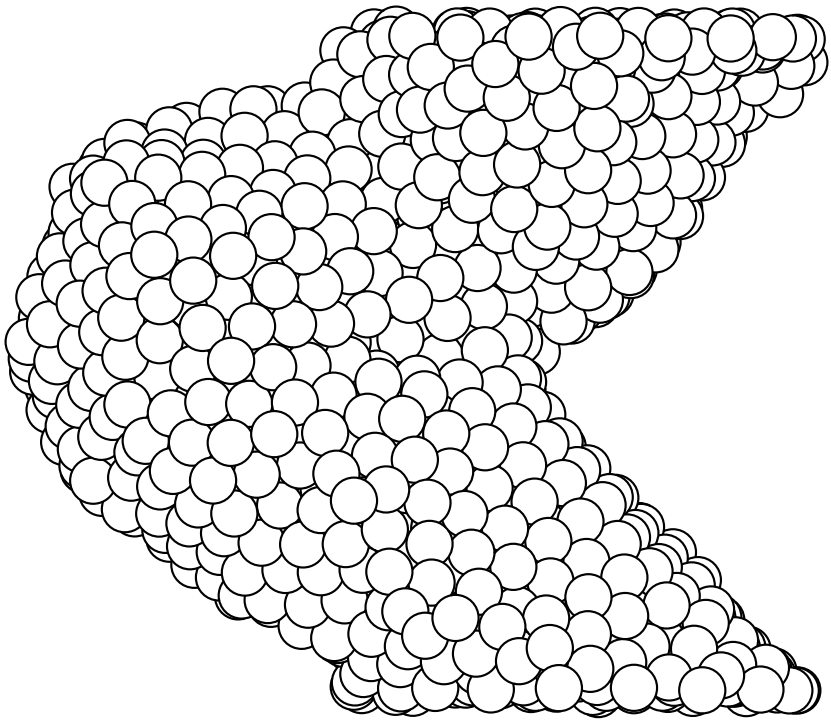
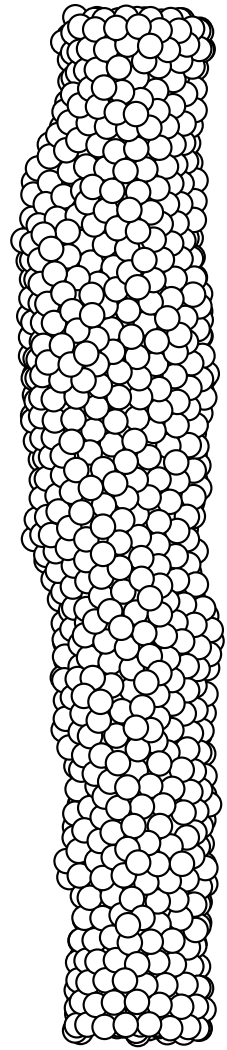
\begin{figure}[tbp]
\caption{ Structural changes in a nanowire shown in Fig. 3(a) and Fig. 1(a):
(a) nanowire deformed at the stress of $6.41$ GPa, (b) the same nanowire
simulated after compression [producing a configuration shown in Fig. 4(a)]
is released.}
\label{fig4}
\end{figure}

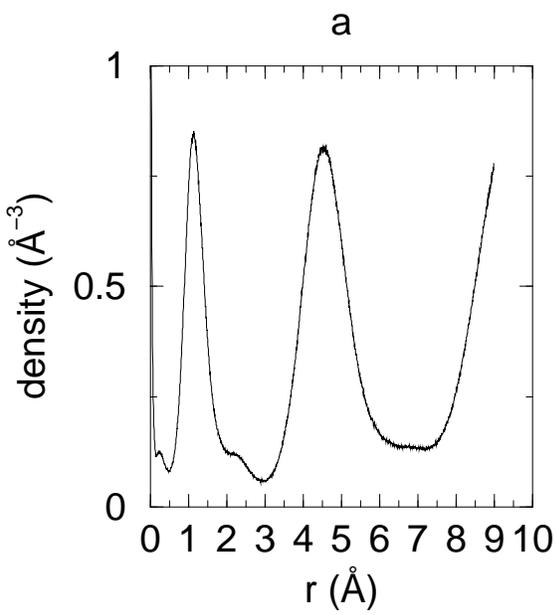
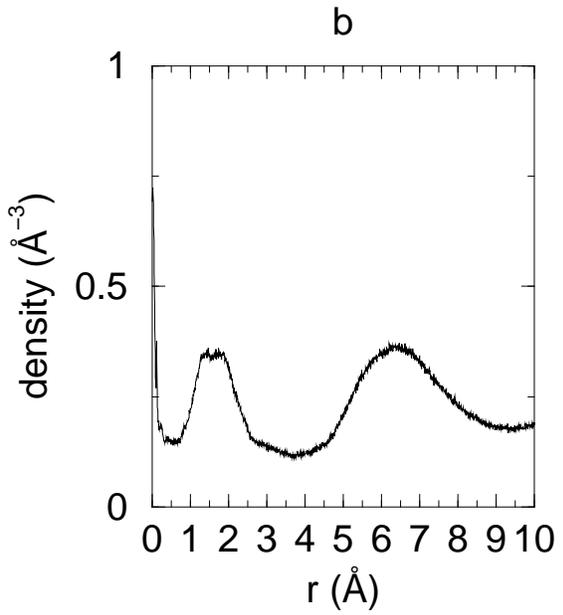
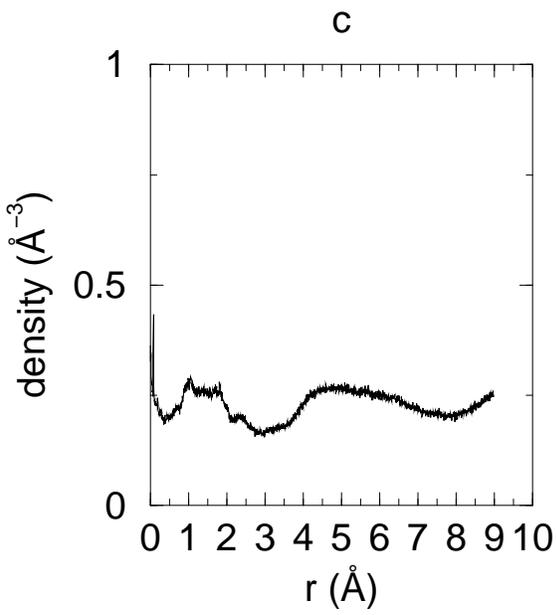
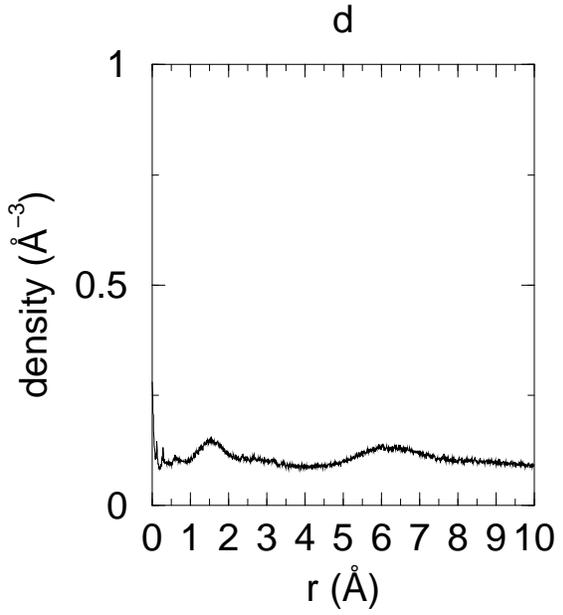
\begin{figure}[tbp]
\caption{The radial density plots for a nanowire with $R_0=0.9$ nm, $L_0=12$
nm, at the stress of: (a) $0$ GPa, (b) $3.2$ GPa, (c) $4.81$ GPa, (d) $6.41$
GPa.}
\label{fig5}
\end{figure}

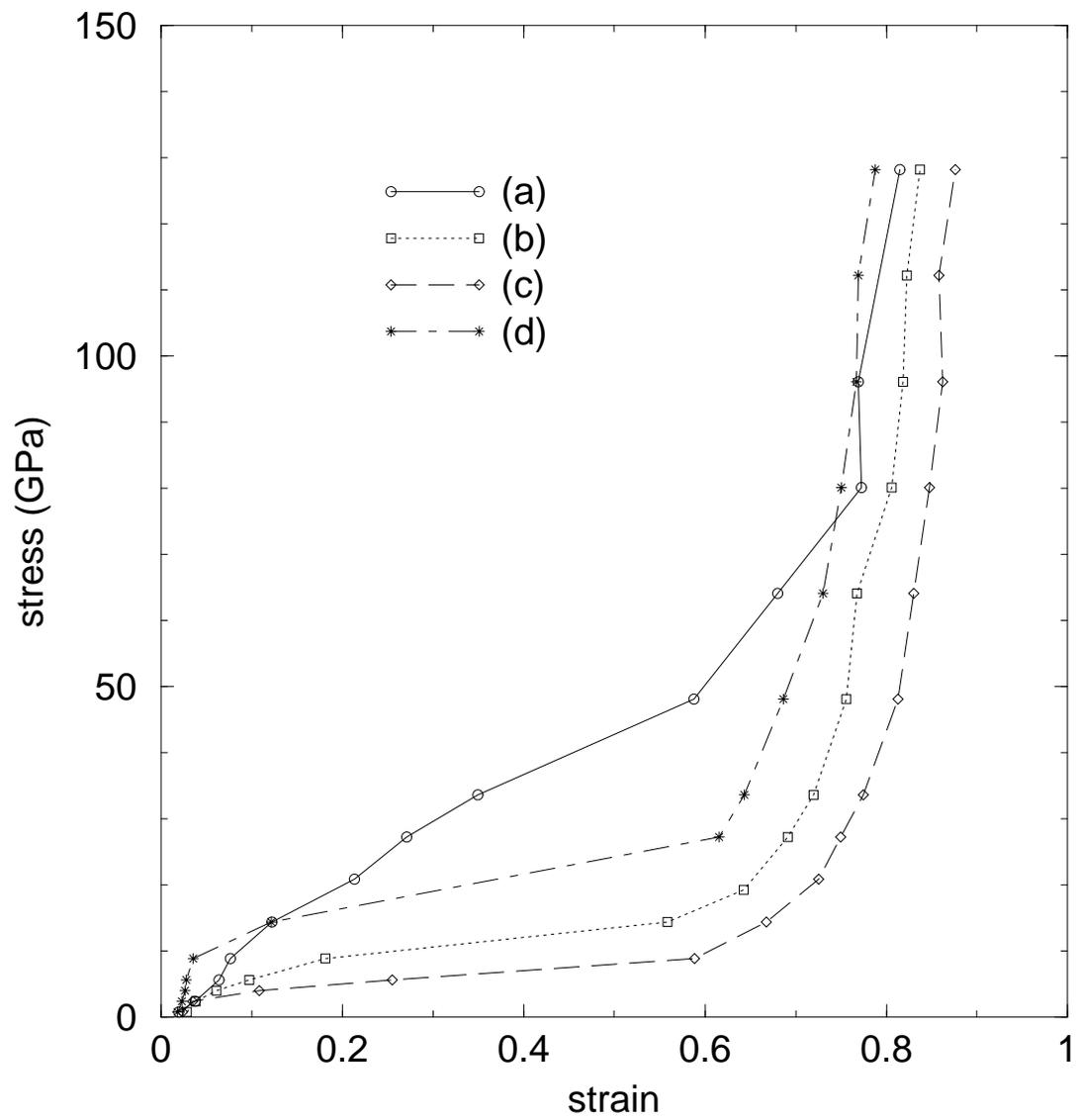
\begin{figure}[tbp]
\caption{Stress versus strain curves: (a) $R_0=0.9$ nm, $L_0=4$ nm, (b) $%
R_0=0.9$ nm, $L_0=8$ nm, (c) $R_0=0.9$ nm, $L_0=12$ nm, (d) $R_0=1.2$ nm, $%
L_0=8$ nm.}
\label{fig6}
\end{figure}

\end{document}